\documentclass[prb,aps,twocolumn,showpacs]{revtex4}

\usepackage{amsmath,amssymb,bm}
\usepackage{graphicx}
\usepackage{longtable}
\usepackage{float}

\usepackage{xcolor}

\begin{document}

\author{Anna Sitek,$^{1,2}$}
 \email{anna.sitek@pwr.wroc.pl}
\author{Andrei Manolescu$^2$}
 \email{manoles@ru.is}
\affiliation{$^1$Institute of Physics, Wroc{\l}aw University of
Technology, 50-370 Wroc{\l}aw, Poland, \\
$^2$School of Science and Engineering, Reykjavik University, Menntavegur 1, 
IS-101 Reykjavik, Iceland}

\title{Dike states in multiple quantum dots}

\begin{abstract}

We present a theoretical study of the collective optical effects which can
occur in groups of three and four quantum dots. We define conditions for
stable subradiant (dark) states, rapidly decaying superradiant states,
and spontaneous trapping of excitation.  Each quantum dot is treated
like a two-level system.  The quantum dots are though realistic, meaning
that they may have different transition energies and dipole moments.
The dots interact via a short-range coupling which allows excitation transfer
across the dots, but conserves the total population.  We calculate
the time evolution of single- and biexciton states using the
Linblad equation.  In the steady state the individual populations of
each dot may have permanent oscillations with frequencies given by the
energy separation between the subradiant eigenstates.

\end{abstract}


\maketitle

\section{Introduction}

Collective optical (superradiant) effects appear in ensembles in which the distances between single 
emitters are much smaller than the radiation wavelength with which they interact (Dicke limit). \cite{Dicke54} 
The spatial dependence of the electromagnetic field within such ensembles is negligible and thus all 
the systems effectively interact with common photon reservoir. This leads to formation of rapidly decaying 
(superradiant) and optically inactive (subradiant) states \cite{Nussenzweig73} and consequently to the 
appearance of a vacuum-induced coherence effect which results in occupation trapping. \cite{Agarwal74}

Although effects resulting from collective coupling of atoms to radiative environment have been known 
for nearly sixty years, are very well described \cite{Dicke54, Nussenzweig73, Agarwal74, Gross82, singh97} 
and have been extensively investigated experimentally, \cite{Skribanowitz73, Pavolini85} 
they still attract much scientific attention. This is caused by the increasing variety of physical 
systems in which these phenomena may be observed, such as quantum dots (QDs), \cite{scheibner07} 
Bose--Einstein condensate, \cite{Baumann10} superconducting qubits, \cite{Fink09, Mlynek12} 
ion Coulomb crystals \cite{Herskind09} or depolaritons.\cite{Kyriienko13}

The investigation of superradiant effects have been initiated and driven to a large extend by the promise 
which the short-living states show for optimization of lasers. \cite{Belyanin03} Recently the scientific 
interest has been focused on the concept of ``superradiant laser'' which allows to increase spectral purity 
of emitted light. \cite{Meiser10a, Meiser10b} The technological realization of such a device presented 
in Ref. \onlinecite{Bohnet12} allows to improve accuracy of atomic clocks \cite{Ludlow08, Jiang11} and 
thus measurements of gravity \cite{Chou10} and fundamental constants. \cite{Fortier07, Blatt08} The 
experimental investigations of the collective effects were mostly restricted to the analysis of the 
superradiant states which appear spontaneously in the cascade emission and manifest themselves as a maximum 
in the intensity or photon emission rate. \cite{Nussenzweig73, Skribanowitz73, Gross76}

Although the observation of the subradiance phenomena was also reported in the atomic ensembles, \cite{Pavolini85} 
as the opposite of the superradiant states, the preparation of subradiant states is much more difficult,
and therefore the possibilities they give were much less investigated experimentally. Recently, the preparation 
of an optically inactive state was reported in a system of superconducting qubits \cite{Filipp11} and in a diatomic 
molecule in an optical lattice. \cite{Takasu12} The advantage of the subradiant states stems from their decoupling 
from the photon environment because of which they do not undergo radiative decoherence and thus may form 
decoherence-free subspaces. \cite{Yamamoto12, Gong08} This makes them useful for quantum information processing 
especially for noiseless encoding of quantum information. \cite{Zanardi97, Zanardi98} The stable states also 
allow to construct a scalable quantum processor, \cite{Petrosyan02} quantum memories, \cite{Kalachev07} 
nonlinear sign-shift gate \cite{Kalachev05} and storing time-bin qubits \cite{Kalachev06} for quantum cryptography. 
Interesting group of Dicke states is formed of single-excitation combinations belonging to the class of 
``W'' states  which have been widely considered for quantum information processing \cite{Zheng06, Agrawal06, Li07, Situ10} 
or optimization of the quantum clock synchronization. \cite{Ben-Av11}

Systems composed of two and more coupled QDs attract much scientific focus due to the richness of their properties 
which pave the way to new technological applications. Already pairs of quantum dots allow for long-time storage of 
quantum information,\cite{pazy01b} conditional optical control of carrier states, \cite{unold05} implementation 
of a two-qubit quantum gate, \cite{biolatti00} optical writing of information on the spin state of the dopant Mn 
atom \cite{goryca09} and construct quantum nanoantennas due to the collective phenomena. \cite{Mokhlespour12} 
Systems of three QDs enable to realize two different kinds of entanglement, \cite{Dur00} teleportation via 
superradiance \cite{Chen05}, CNOT gates\cite{Kestner11} and the control of spin blockade. \cite{Busl10} Moreover, 
in these systems the collective transport effects (electronic Dicke or Kondo-Dicke effect) may be 
realized \cite{Trocha08, Vernek10} and lead to the enhancement of thermoelectric efficiency. \cite{Wang13} 
Arrays of QDs allow to reduce the effect of pure dephasing on quantum information encoded in excitonic 
states. \cite{Grodecka06}

In this paper we analyze the collective optical effects in ensembles of three and four QDs. Compared to 
double QDs in which only one optically inactive state may be realized, \cite{Nussenzweig73} ensembles of three 
and more two-level systems allow to realize many stable states which occur at different exciton occupations of 
the single emitters. Although the superradiant effects are well described in systems of identical atoms the 
description of such phenomena in QD ensembles requires taking into account characteristic for those system 
properties which distinguish them from natural ones. Therefore we include in our model the fundamental energy 
mismatches, different dipole moments of single dots and coupling which induces excitation transfer between single 
emitters, but conserves the total population of the ensemble. In our previous works concerning double QDs it has 
been shown that the collective optical effects are extremely sensitive to inhomogeneity of the fundamental transition 
energy which leads to the decay of the exciton occupation for the energy splitting much below the present technological 
feasibility. This decoherence effect may be strongly reduced by sufficiently strong coupling between 
the dots \cite{sitek07a, sitek09a, sitek09b} and fully overcome in double QDs with different decay rates. \cite{sitek12} 
In this paper we extend the results of two QDs and specify conditions which allow to take advantage of the superradiant 
phenomena in fully inhomogeneous QD systems. We analyze the dynamics of one electron-hole pairs and biexcitons. We 
show how to adjust the system parameters in such a way that an arbitrary dark single-exciton combination may be blocked 
in a multiple QD, we also specify the conditions which allow to trap two excitons and prepare a system in a biexcitonic 
state which allows to recombine only one electron-hole pair. Due to existence of two or more subradiant single-exciton 
eigenstates and coupling between the dots the occupation of individual dots oscillates while the population of the whole 
ensemble remains stable, we show that the oscillation amplitudes may be strongly reduced if the system is initially prepared 
in a biexciton state.

The paper is organized as follows. In Sec. \ref{sec:system} we describe the system under study, define its model 
and describe a method used to study the system evolution, in Sec. \ref{sec:results} we present and discuss
our results, we conclude the paper in Sec. \ref{sec:conclusions}.

\section{The system}
\label{sec:system}

The investigated system consists of $N$ $(N=3,4)$ quantum dots (QDs) in which only the ground-level exciton 
states with fixed spin polarizations are taken into consideration. Due to the strong Coulomb coupling and 
absence of the external electric fields we may restrict the discussion to `spatially direct' excitonic 
states, i.\ e., states with electron-hole pairs residing in the same QD which in these conditions have much 
lower energy than the `dissociated' states referring to excitons formed of carriers residing in two 
different dots. \cite{szafran05,szafran08} These assumptions allow us to treat every QD as a two-level 
system which may be either empty or contain an exciton and thus describe the set of $N$ QDs as 
an $2^N$-level system, with $|0\rangle$ denoting the ground (or ``vacuum'') state in which all $N$ dots 
are empty, single-exciton states $|i\rangle$, corresponding to one exciton localized in the $i$th QD, 
biexcitonic states $|ij\rangle$ referring to electron-hole pairs residing in the $i$th and $j$th QDs, 
states $|ijk\rangle$ with three QDs, $i$th, $j$th, and $k$th ($1\leqslant i,j,k\leqslant N$), 
occupied with an exciton \textit{etc}.

Present manufacturing technology does not allow to produce on demand systems of QDs with identical 
fundamental transition energies, therefore we assume different electron-hole binding energies of
each dot defined as 
\begin{eqnarray*}
 E_{i} = E + \Delta_{i},
\end{eqnarray*}
where $E$ is the average transition energy and $\Delta_{i}=\alpha_{i}\Delta$ is the energy mismatch 
of the $i$th QD. We impose $\sum_{i}\alpha_{i}=0$ and $\sum_{i}\alpha^{2}_{i}=1$, such that $\Delta^{2}$ 
is the mean square variation of the transition energies.

We analyze the system in a 'rotating frame' defined by the evolution operator
\begin{eqnarray*}
U = \exp\left[-\frac{i}{\hbar}\left(E\sum\limits^{N}_{i}\sigma^{(i)}_{+}\sigma^{(i)}_{-}  +  H_{\mathrm{rad}}\right)t\right],
\end{eqnarray*}
where $\sigma^{(i)}_{-}=\left(\sigma^{(i)}_{+}\right)^\dagger
=|0\rangle\!\langle i| + \sum_{j}|j\rangle\!\langle ij| + \sum_{jk}|jk\rangle\!\langle ijk| + \cdots$ 
are the annihilation (creation) operators for the exciton in the $i$th QD, respectively and 
$H_{\mathrm{rad}}=\sum_{\bm{k}\lambda}\hbar\omega_{\bm{k}}b^{\dagger}_{\bm{k}\lambda}b_{\bm{k}\lambda}$ 
is the standard free photon Hamiltonian with operators $b^{\dagger}_{\bm{k}\lambda}$ and $b_{\bm{k}\lambda}$  
creating and annihilating radiation modes with wave vector $\bm{k}$ and polarization $\lambda$, while
$\omega_{\bm{k}}$ is the corresponding frequency.

In this frame the Hamiltonian of the system is
\begin{eqnarray*}
H=H_{\mathrm{S}}+H_{\mathrm{S-rad}}.
\end{eqnarray*}
The first term describes electron-hole pairs residing in the QD system. We assume that the ground state 
$|0\rangle$ corresponds to the zero energy level, so the excitonic Hamiltonian is
\begin{eqnarray}
\label{hamiltonianS}
H_{\mathrm{S}} &=& \sum\limits_{i=1}\Delta_{i}\sigma^{(i)}_{+}\sigma^{(i)}_{-}
+ \sum\limits_{i,j=1}B_{ij}|ij\rangle\!\langle ij|    \nonumber \\
&+& \sum\limits_{i,j,k=1}B_{ijk}|ijk\rangle\!\langle ijk|+ \cdots   \nonumber \\
&+& \sum\limits_{i,j=1}V_{ij}\sigma^{(i)}_{+}\sigma^{(j)}_{-},
\end{eqnarray}
where $B_{ij}$ are biexcitonic shifts due to the interaction of static dipole moments of $i$th 
and $j$th QD, $B_{ijk}$ is a deviation of energy caused by interaction of the dipole moments
of three dots, \textit{etc}.

In ensembles of QDs one may distinguish two types of interaction between the emitters: dipole 
(F{\"o}rster) coupling which in the leading order decays as $1/r_{ij}^3$ with the QD 
separation \cite{danckwerts06, singh07, singh12} and short-range coupling resulting form a combination of 
tunneling (wave function overlap) and Coulomb correlations. Both types of interaction induce excitation 
transfer between the emitters but conserve the total exciton occupation of the ensemble. The short-range 
couplings allow to rebuilt enhanced emission in the energetically inhomogeneous ensembles while in the case 
of dipole interaction a similar effect is achieved if the coupling is enhanced artificially by a factor 
of $400$. \cite{kozub12} To overcome the destructive effect of the transition energy mismatch the coupling 
between the dots must be of the order of the energy splitting, i.\ e., $1$ meV for technologically feasible 
systems. \cite{sitek07a, sitek09a, sitek09b} For a planar QD arrangement the distance between emitters is 
about $30$ nm (the average value for the sample studied in Ref. \onlinecite{scheibner07}), and for such a 
distance the F\"orster coupling drops to about $1$ $\mu$eV, which is not sufficient to stabilize collective 
effects.

The F{\"o}rster coupling reaches a fraction of meV for QDs separated by only a few nanometers,\cite{lovett03} 
which can be achieved in vertically stacked QD systems, and where indeed it can stabilize the collective 
effects.\cite{sitek07a,sitek12} In the present paper we describe a planar system and therefore we prefer
to consider a short-range coupling, for which we choose an exponential model, 
$V_{ij} = V_{0}\exp\left[-r_{ij}/r_{0}\right]$, where the subscripts $i$ and $j$ refer to $i$th and $j$th QD, 
respectively, $V_{0}$ is a constant amplitude, $r_{ij}$ is the distance between the QDs and $r_{0}$ the spatial 
range of the interaction. Remarkably, this simple model reproduced well recent experimental results.\cite{kozub12}

The eigenstates of the Hamiltonian (\ref{hamiltonianS}) do not mix quantities associated with different
exciton numbers, i.\ e. the eigenstates of the system are superpositions of the basis states restricted to a 
particular exciton number.

The second term of the Hamiltonian accounts for coupling between the QD system and quantum electromagnetic field
\begin{eqnarray}
\label{hamiltonianSrad}
H_{\mathrm{S-rad}}=\sum_{j=1}^{N}\sum_{\bm{k}\lambda}
\sigma^{(j)}_{-}g^{(j)}_{\bm{k}\lambda}
e^{-i\left(\frac{E}{\hbar}-\omega_{\bm{k}}\right)t}b^{\dagger}_{\bm{k}\lambda} 
+\mathrm{H.\ c}.,
\end{eqnarray}
where 
\begin{displaymath}
g^{(j)}_{\bm{k}\lambda}=i\bm{d}_{j}\cdotp\!\hat{e}_{\lambda}(\bm{k})
\sqrt{\frac{\hbar\omega_{\bm{k}}}{2\epsilon_{0}\epsilon_{r}v}}
\end{displaymath} 
is a coupling constant for the $j$th QD. Here $\bm{d}^{(j)}$ is the inter-band dipole moment for the 
$j$th QD, $\hat{e}_{\lambda}(\bm{k})$ is the unit polarization vector of the photon mode with 
polarization $\lambda$, $\epsilon_0$ is the vacuum dielectric constant, $\epsilon_r$ is the relative 
dielectric constant and $v$ is the normalization volume. We restrict our investigations to wide-gap 
semiconductors with electron-hole binding energies of the order of 1 eV which allows us to describe
the photon modes within the zero-temperature approximation at any reasonable temperature.

To describe the evolution of the carrier subsystem we use an equation of motion for the reduced density 
operator in the Markov approximation. In the 'rotating frame' it takes a form
\begin{eqnarray*}
 \dot{\rho}=-\frac{i}{\hbar}\left[H_{\mathrm{S}},\rho\right]+\mathcal{L}_{\mathrm{rad}}[\rho],
\end{eqnarray*}
where $\rho$ is a reduced density matrix of the exciton subsystem and $\mathcal{L}_{\mathrm{rad}}$ is
a Lindblad dissipator
\begin{eqnarray*}
\mathcal{L}_{\mathrm{rad}}[\rho] 
= \sum_{i,j=1}^{N}\Gamma_{ij}\left[\sigma^{(i)}_{-}\rho\sigma^{(j)}_{+}
-\frac{1}{2}\left\{\sigma^{(j)}_{+}\sigma^{(i)}_{-},\rho\right\}\right],
\label{superLindblad}
\end{eqnarray*}
where 
\begin{eqnarray}
\label{Gamma}
\Gamma_{ij}=\Gamma^{*}_{ij}=\frac{E^{3}}{3\pi\epsilon_{0} \epsilon_{r}\hbar^{4}c^3}
\bm{d}_{i}\!\cdot\!\bm{d}^{*}_{j}
\end{eqnarray}
with $c$ being the speed of light.
Since for $i = j$ the Eq. (\ref{Gamma}) describes the spontaneous decay rates of single QDs, \cite{scully} 
the mixed (off-diagonal) decay rates (\ref{Gamma}) may be expressed in terms of single QD quantities, 
\begin{eqnarray*}
\Gamma_{ij}=\Gamma_{ji}^{*}
=\sqrt{\Gamma_{ii}\Gamma_{jj}}\hat{\bm{d}}_{i}\!\cdot\!\hat{\bm{d}}^{*}_{j},
\end{eqnarray*}
where $\hat{\bm{d}}_{i}=\bm{d}_{i}/d_{i}$ and $\hat{\bm{d}}_{i}\cdot\hat{\bm{d}}^*_{j} \approx e^{i\eta} (1-\theta^2_{ij}/2)$,
here $\eta$ is an irrelevant phase and $\theta_{ij}$ is a small angle between the dipole moments which depends  
of light-hole admixture. \cite{sitek12}

In the numerical simulations we assume constant energy mismatches with the parameters $\alpha_{1}=2/\sqrt{56}$, 
$\alpha_{2}=4/\sqrt{56}$, $\alpha_{3}=-6/\sqrt{56}$, $\alpha_{4}=0$ and $\Delta = 1$ meV (except for the 
Fig. \ref{fig:3QD_angle}). For the coupling 
amplitudes we take $V_{0} = 5$ meV and $r_{0} = 15$ nm. In Sec. \ref{idealsingle}, \ref{coupledsingle} 
and \ref{sec:biexciton}  we 
assume parallel dipole moments, since the effect of the light-hole admixture in the absence of external 
electric fields is negligible. \cite{sitek12}

\section{Results}
\label{sec:results}

\begin{figure}[t]
\centering
\includegraphics[scale=0.1]{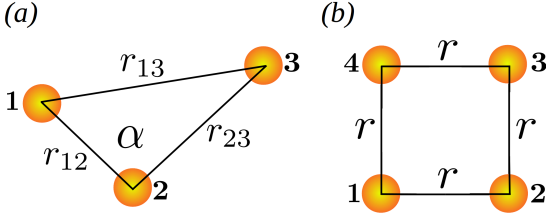}
\caption{The planar arrangement of the three (a) and four (b) QD ensembles.}
\label{fig:34QD}
\end{figure}

Below we present an analysis of the collective effects which occur in multiple QDs. We define 
the Dicke states for an arbitrary number of emitters and perform numerical simulations 
for ensembles of three and four QDs. In Fig. \ref{fig:34QD} we illustrate the numbering of the QDs
and their spatial arrangement. In Sec. \ref{idealsingle} and \ref{coupledsingle} we focus on 
single-exciton states. In Sec. \ref{idealsingle} we show the evolution of uncoupled systems
with identical fundamental transitions and parallel dipole moments of different amplitudes. 
Then, in Sec. \ref{coupledsingle} we analyze the same effects in a system composed of three coupled
energetically inhomogeneous dots. The dynamics of biexciton states is presented in Sec. \ref{sec:biexciton}.
In Sec. \ref{pinjunction} we show possibilities of controlling the exciton occupation given by a p-i-n 
junction.

\subsection{Single-exciton states of ideal quantum dots}
\label{idealsingle}

The collective effects were described for the first time in ensembles of uncoupled identical atoms where 
all the emitters have the same transition energies and dipole moments. \cite{Dicke54}  Such optical effects, 
resulting from the interaction in the Dicke limit, are also present in uncoupled systems, even with
different dipole moments, if all the emitters have identical transition energies. For the purpose of this paper 
we define such systems as ideal QDs.

The coupling of excitons to their radiative environment described in the Dicke limit by the 
Hamiltonian (\ref{hamiltonianSrad}) and the Fermi's Golden Rule, according to which the probability 
of releasing a photon through a transition from the initial to the final states is 
$P\sim|\langle\mathrm{final}|H_\mathrm{S-rad} |\mathrm{initial}\rangle|^2$,
allow to define rapidly decaying (superradiant) and optically inactive (subradiant) states, both also known as 
Dicke states. By definition the superradiant initial state  ($|\mathrm{SUPER}\rangle$) corresponds to the maximum 
transition probability, whereas the subradiant states ($|\mathrm{SUB}\rangle$) refer to the vanishing probability. 
Due to the decoupling from the photon reservoir these are dark, optically inactive states. In the weak excitation 
limit, i.\ e., for a single excitation in the system from which the sample may decay only to the ground state 
($|0\rangle$), the proportionality $g^{(j)}_{\bm{k}\lambda}\sim\sqrt{\Gamma_{jj}}$ allows to write the short-living 
states in the form
\begin{eqnarray}
\label{bright}
 |\mathrm{SUPER}\rangle = \frac{\sum\limits^{N}_{i=1}\sqrt{\Gamma_{ii}}|i\rangle}{\sqrt{\sum\limits^{N}_{i=1}\Gamma_{ii}}},
\end{eqnarray}
and express the stable superpositions as 
\begin{eqnarray}
\label{dark}
|\mathrm{SUB}\rangle =\frac{\sum\limits^{N}_{i=1}a_{i}\sqrt{\prod\limits^{N}_{j=1}\frac{\Gamma_{jj}}{\Gamma_{ii}}}|i\rangle}
{\sqrt{\sum\limits^{N}_{i=1}|a_{i}|^2\prod\limits^{N}_{j=1}\frac{\Gamma_{jj}}{\Gamma_{ii}}}},
\end{eqnarray}
where the coefficients $a_{i}$ satisfy $\sum_{i}a_{i}=0$. Irrespective of the number of emitters there 
is only one superradiant state in the system of a particular number of QDs, whereas the only structure 
which realizes just one dark state is a double quantum dot. The systems of three and more QDs allow to 
realize an arbitrary number of dark states of the form ($\ref{dark}$) since there are may combinations 
of the parameters $a_{i}$ for which the transition matrix element 
$\langle 0|H_\mathrm{S-rad} |\mathrm{SUB}\rangle=0$.

\begin{figure}[t]
\centering
\includegraphics[scale=0.7]{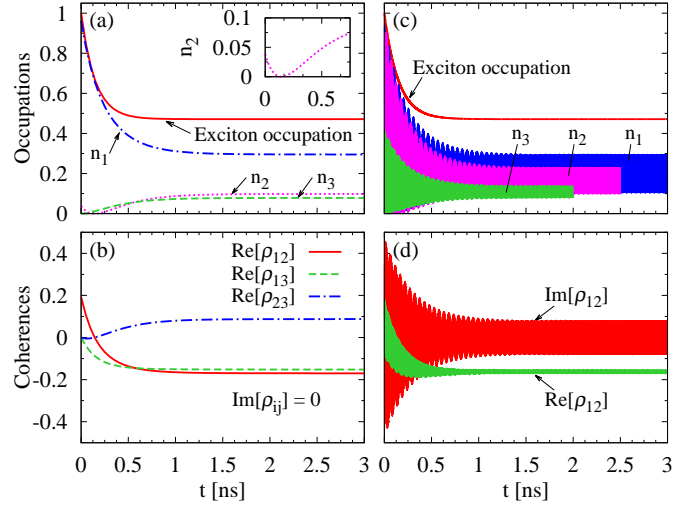}
\caption{Exciton occupations [(a) and (c)] and coherences [(b) and (d)] of a system of 
three QDs prepared initially in a bright state $\left(5|1\rangle+|2\rangle\right)/\sqrt{26}$ for 
$\Gamma_{11}=2.44$ ns$^{-1}$, $\Gamma_{22}=3.31$ ns$^{-1}$ and $\Gamma_{33}=1$ ns$^{-1}$. 
Figs. (a) and (b) show the results for uncoupled QDs ($V_{ij}=0$) with identical transition energies 
($\Delta=0$), while Figs. (c) and (d) refer to energetically inhomogeneous ($\Delta\neq 0$) and coupled
($V_{ij}\neq 0$) system arranged in a lateral array of an equidistant triangle shape with the side length 
$r=30$ nm. The inset to Fig. (a) shows the initial evolution of the exciton occupation of the QD number 2.}
\label{fig:3QD_comparison}
\end{figure}

The consequence of the co-existence of rapidly decaying and stable states is the effect of spontaneous 
trapping of excitation. \cite{Agarwal74, sitek12} An arbitrary single-exciton state 
$|\mathrm{s}\rangle=\sum_{i}c_{i}|i\rangle$, where $|c_{i}|^2$ is the localization probability of the 
exciton on the $i$th QD and $\sum_{i}|c_{i}|^2=1$, may be expressed as a combination of the superradiant 
state (\ref{bright}) and a dark state of type (\ref{dark}),
\begin{eqnarray}
\label{single}
 |\mathrm{s}\rangle &=& \frac{\left(\sum\limits^{N}_{i=1}c_{i}\sqrt{\Gamma_{ii}}\right)
 }{\sqrt{\sum\limits^{N}_{i=1}\Gamma_{ii}}}|\mathrm{SUPER}\rangle \nonumber \\
 &+&
\sqrt{1-\frac{\left(\sum\limits^{N}_{i=1}c_{i}\sqrt{\Gamma_{ii}}\right)^2
 }{\sum\limits^{N}_{i=1}\Gamma_{ii}}} 
 |\mathrm{SUB}\rangle_{\mathrm{s}},
\end{eqnarray}
where the dark state is
\begin{eqnarray}
\label{dark2}
  |\mathrm{SUB}\rangle_{\mathrm{s}} = \frac{\sum\limits^{N}_{i=1}c_{i}
  \left[\left(\sum\limits^{N}_{j\neq i}\Gamma_{jj}\right)|i\rangle-\sum\limits^{N}_{j\neq i}\sqrt{\Gamma_{ii}\Gamma_{jj}}|j\rangle\right]}
  {\sqrt{\left(\sum\limits^{N}_{i=1}\Gamma_{ii}\right)
  \left[\sum\limits^{N}_{i=1}\Gamma_{ii}-\left(\sum\limits^{N}_{i=1}c_{i}\sqrt{\Gamma_{ii}}\right)^2\right]}}.
\end{eqnarray}
The derivation of the Eqs. (\ref{single}) and (\ref{dark2}) is done in the the Appendix.
The collective coupling to the radiative surrounding induces emission only from the superradiant state and 
thus the fraction of excitation initially spanned in the dark state, 
$1-\left(\sum\limits^{N}_{i=1}c_{i}\sqrt{\Gamma_{ii}}/\sqrt{\sum\limits^{N}_{i=1}\Gamma_{ii}}\right)^2$, 
remains unaffected. Since the only single-exciton state which decays totally is the superradiant state, 
we define a state $|\mathrm{s}\rangle$ as being bright if its superradiant contribution does not vanish, i.\ e., 
if a system prepared in that state partially recombines and only a part of the initial exciton occupation 
remains trapped.

In Fig. \ref{fig:3QD_comparison} we show the dynamics of a
single-excitation induced by a common photon reservoir in an ideal system
of three uncoupled ($V_{ij} = 0$) QDs with equal electron-hole binding
energies ($\Delta = 0$) and parallel dipole moments ($\theta_{ij}=0$)
of different magnitudes.  In Fig. \ref{fig:3QD_comparison}(a) we show
the evolution of the exciton occupations of a system prepared initially
in a bright state $\left(5|1\rangle+|2\rangle\right)/\sqrt{26}$ and in
Fig. \ref{fig:3QD_comparison}(b) we show the corresponding coherences. As
expected, the coupling to the photon reservoir spans the excitation
into the sub- and superradiant states according to Eq. (\ref{single})
and induces emission only from the short-living state
\begin{eqnarray*}
 \frac{\sqrt{\Gamma_{11}}|1\rangle + \sqrt{\Gamma_{22}}|2\rangle + \sqrt{\Gamma_{33}}|3\rangle}
 {\sqrt{\Gamma_{11}+\Gamma_{22}+\Gamma_{33}}},
\end{eqnarray*}
[Fig. \ref{fig:3QD_comparison}(a)]. The excitation dynamics takes place until occupations of all dots stabilize 
at certain levels corresponding to the dark state defined in the Eq. (\ref{dark2}) which confirms that the 
state (\ref{dark2}) is indeed unaffected by the photon reservoir and, after the decay of the superradiant state, 
neither the total exciton occupation nor occupations of single dots ($n_{1,2,3}$) change due to radiative environment.
The emission from the above state induces decay of the total exciton 
occupation and excitation transfer which results in the redistribution of the occupations of single dots.
Since all of the localized single-exciton states $|i\rangle$ contribute to the superradiant state (\ref{bright}), 
the collective coupling spans the initial excitation in all of the dots even if some of them were initially empty. 
Therefore the population of initially unoccupied systems builds up spontaneously [$n_3$, green-dashed line 
in Fig. \ref{fig:3QD_comparison}(a)]. If the initial occupation of one of the dots is relatively small while 
the spontaneous decay rate from that system is sufficiently strong then the exciton occupation of that dot may 
vanish at same point and then restore due to the excitation transfer [magenta-dotted line 
in Fig. \ref{fig:3QD_comparison}(a) and the inset to Fig. \ref{fig:3QD_comparison}(a)]. 
During the emission process also the evolution of the off-diagonal density matrix elements is 
observed, the coherences related to the initially populated dots decay, while those corresponding to initially empty 
systems build up spontaneously due to the increasing occupations of those dots. When exciton dynamics in the 
system reaches population distribution corresponding to the optically inactive state also the off-diagonal 
density matrix elements stabilize at a certain non-zero level [Fig. \ref{fig:3QD_comparison}(b)], defined by 
the dark contribution to the initial state.

\subsection{Single-exciton states for inhomogeneous quantum dots}
\label{coupledsingle}

Technologically feasible QDs forming multiple structures differ both in fundamental transition energies 
($\Delta\neq 0$) and dipole moments ($\bm{d}_i\neq\bm{d}_j$), and are coupled with each other ($V_{ij}\neq 0$). 
As shown in the previous section and, for a double 
QD in Ref. ~\onlinecite{sitek12},  the superradiant character of the evolution of one exciton is present in 
the ideal systems ($\Delta=0$), but with parallel dipole moments ($\theta_{ij}=0$). The collective evolution is very 
sensitive to the energy mismatches and is destroyed in ensembles with energy splittings of the order of the 
transition line width. \cite{sitek12,sitek07a,sitek09a,sitek09b} In such systems the localized eigenstates 
corresponding to different energies cannot form delocalized superpositions which would also be the system 
eigenstates. This destructive effect may be overcome by coupling between the dots ($V_{ij}$) which delocalizes 
the system eigenstates and different dipole moments allowing the superradiant state to be a non-symmetric 
superposition of the localized states $|i\rangle$ [Eq.(\ref{bright})].

The single-exciton eigenstates of the system depend on the energy mismatches and coupling between the dots, 
while the Dike states are defined by the interplay of decay rates [Eqs. (\ref{bright}) and (\ref{dark})]. 
If the single-exciton decay rates 
[Eq. (\ref{Gamma}) for $i=j$] are adjusted in such a way that the superradiant state (\ref{bright}) corresponds 
to one of the system eigenstates, then the inhomogeneous ensemble of QDs interacts with its radiative environment 
in the ``collective regime'' i.\ e. allows many effects typically present only in systems with identical electron-hole 
binding energies. The amplitudes $c_{i}$ of a single-exciton 
state orthogonal to the superradiant state (\ref{bright}) must satisfy the equation $\sum_{i}c_{i}\sqrt{\Gamma_{ii}}=0$ 
which implies the condition $\langle 0|H_{\mathrm{S-rad}}|\mathrm{SUB}\rangle=0$ defining a subradiant state. 
Therefore, if one of the eigenstates has a superradiant character then the other eigenstates of a system are optically 
inactive and thus defying the ``collective regime'' requires only specifying the superradiant eigenstate.

In Figs. \ref{fig:3QD_comparison}(c) and \ref{fig:3QD_comparison}(d) we show the evolution of a realistic group of three 
QDs placed in the corners of an equilateral triangle, we assume non-equal fundamental transition energies 
($\Delta = 1$ meV), non-vanishing coupling between the systems ($V_{ij}\neq 0$) and parallel dipole moments ($\theta_{ij}=0$). 
We compare 
the results obtained for an inhomogeneous system coupled to the photon reservoir in the ``collective regime'' to the 
ideal case presented in Figs. \ref{fig:3QD_comparison}(a) and \ref{fig:3QD_comparison}(b), where the decay rates of individual 
dots take the same values as in Figs. \ref{fig:3QD_comparison}(c) and \ref{fig:3QD_comparison}(d). As can be seen in 
Fig. \ref{fig:3QD_comparison}(c),  coupled ensembles with energy mismatches of the order of meV 
allow to trap the same fraction of excitation as ideal dots with the same decay rates [red-solid lines in 
Figs. \ref{fig:3QD_comparison}(a) and \ref{fig:3QD_comparison}(c)]. Here, as in the ideal case, any 
single-exciton eigenstate may be decomposed into sub- and superradiant component according to Eq. (\ref{single}) and 
also in this case the superradiant state is the only state which decays totally. For an arbitrary set of decay rates, 
which do not correspond to the superradiant eigenstate, the exciton occupation of a system prepared initially 
in a state of the form (\ref{dark}) is quenched and the decay of the state (\ref{bright}) is slowed down compared to 
the ``collective regime''.

\begin{figure}[t]
\centering
\includegraphics[scale=0.7]{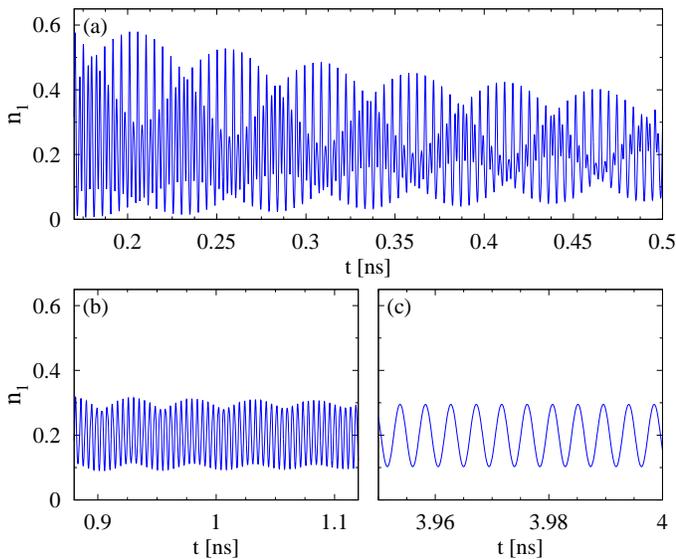}
\caption{The exciton occupation of the QD number 1 ($n_{1}$) 
shown by the blue line in Fig. \ref{fig:3QD_comparison}(c).
The time evolution is shown for three different time 
intervals at picosecond resolution scale. 
}
\label{fig:3QD_oscill}
\end{figure}

Although sub- and superradiant states may exist in the appropriately designed realistic systems, the dynamics 
of individual QD occupations differs considerably from the discussed in the previous section ideal case, 
when the dots interact only through the common radiative reservoir. 
The situation changes when the dots communicate with each other via short-range coupling which induces excitation 
transfer between dots and thus oscillations in the evolution of single QD populations. In a double QD 
system the oscillation amplitudes decrease and the occupations stabilize at levels corresponding to the only one 
dark state of the system. \cite{sitek12} In the multiple QDs composed of three or more emitters oscillations of
the individual dot populations never vanish [Figs. \ref{fig:3QD_comparison}(c), \ref{fig:3QD_comparison}(d) and 
Fig. \ref{fig:3QD_oscill}]. The excitation is trapped in the system because of the existence of dark states, which 
in such systems may be realized by many different amplitude combinations and thus also the final population number 
may be realized in various ways.

In Fig. \ref{fig:3QD_oscill} we zoom the exciton occupation $n_{1}$ shown in 
Fig. \ref{fig:3QD_comparison}(c) (blue line).
As can be seen in Fig. \ref{fig:3QD_oscill}(a) the initial oscillation pattern is relatively complicated which 
is caused by the existence of three energy gaps defined by the differences between the superradiant eigenstate 
and the subradiant ones, $|E_{\mathrm{SUPER}}-E_{\mathrm{SUB}_{1(2)}}|$, and by the energy splitting between 
the two subradiant eigenstates, $|E_{\mathrm{SUB}_{1}}-E_{\mathrm{SUB}_{2}}|$. 
The period of the envelope oscillations is a multiplication of the corresponding three periods,  
i.\ e. $T_{1(2)}=h/|E_{\mathrm{SUPER}}-E_{\mathrm{SUB}_{1(2)}}|)$
and $T=h/|E_{\mathrm{SUB}_{1}}-E_{\mathrm{SUB}_{1}}|$. 
The period $T$ itself defines the fine oscillations of 
the occupation. Due to the emission process, the superradiant contribution attenuates and interference pattern 
simplifies [Fig. \ref{fig:3QD_oscill}(b)]. Finally, when the short-living state decays and the total exciton 
occupation becomes trapped, the evolution of the single dot occupation shows single-mode pattern which repeats 
with the time $T$ and with amplitude depending of the initial occupation of individual dots
[Fig. \ref{fig:3QD_oscill}(c)].

\begin{figure}[t]
\centering
\includegraphics[scale=0.7]{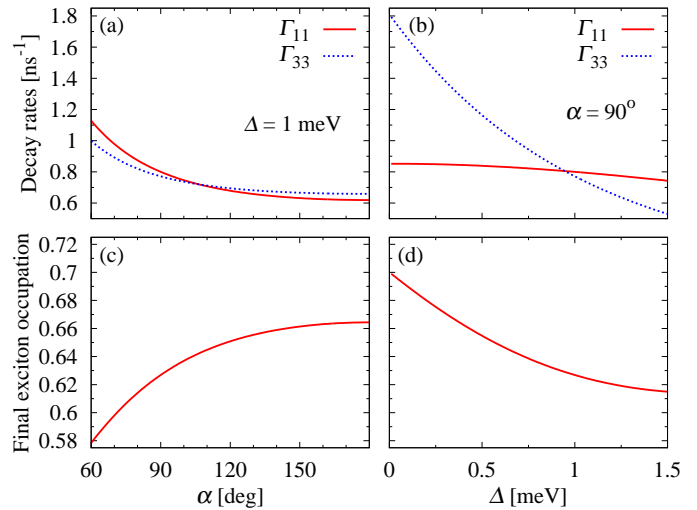}
\caption{Decay rate (in the ``collective regime'') dependence on the spatial arrangement of 
a triple QD i.\ e., on the angle $\alpha$ defined in Fig. \ref{fig:34QD} (a), 
on the energy mismatch (b) and the corresponding steady state occupations for the 
initial state $(5|1\rangle+|2\rangle)/\sqrt{26}$ (c) and (d), respectively. For all Figs. 
we assume $r_{12} = 30 $ nm and $r_{23} = 20 $ nm and a constant 
decay rate of the QD number 2 ($\Gamma_{22} = 2$ ns$^{-1}$).}
\label{fig:3QD_angle}
\end{figure}

The system of three and more QDs allows many different planar arrangements of the emitters. Since the coupling 
amplitudes ($V_{ij}$) depend on the distances between emitters, the eigenstates of the system, and thus the 
decay rates for which the ensemble interacts collectively with its radiative environment, also depend on 
the arrangement of the dots. 
Using the geometry design defined in Fig. \ref{fig:34QD}(a) we  calculate the 
dependence of the decay rates $\Gamma_{11}$ and $\Gamma_{33}$ on the spatial arrangements of the system. 
We assume constant distances $r_{12}$ and $r_{23}$ and thus constant values of the coupling amplitudes $V_{12}$ and $V_{23}$
and change the angle $\alpha$ from $60$ degrees to linear design, i.\ e., we increase the distance between dots $1$ 
and $3$. As seen in Fig. \ref{fig:3QD_angle}(a) the values of the decay rates necessary 
to form the ``collective regime'' slightly decrease with increasing angle $\alpha$.
The two similarly decreasing decay rates (while the third one is constant), according to Eq. (\ref{single})
lead to increasing steady state (final) exciton occupation which is shown in Fig. \ref{fig:3QD_angle}(c).
If the coupling between two out of three QDs is much stronger than coupling of that 
dots with the third one (e.g. $V_{13}\gg V_{12},V_{23}$, $r_{13}\ll r_{12},r_{23}$) then one of the system 
eigenstates has large contribution from the localized state associated with the weakly coupled dot (number $2$)
and to achieve ``collective 
regime'' the decay rate $\Gamma_{22}$ must be much smaller from $\Gamma_{11}$ and $\Gamma_{33}$. In the limiting 
case of vanishing couplings $ V_{12}$ and $V_{23}$ the localized state $|2\rangle$
becomes the system eigenstate and the corresponding decay rate ($\Gamma_{22}$) vanishes. Consequently the pair of 
coupled dots acts as a double QD while the third dot does not contribute to the evolution.

In Fig. \ref{fig:3QD_angle}(b) we show the dependence of the decay rates forming the ``collective regime'' 
on the energy mismatch. Both calculated decay rates decrease with increasing energy separation, but one 
changes slowly while the second decays fast, this leads to decreasing (Eq. \ref{single}) steady state 
occupation shown in Fig. \ref{fig:3QD_angle}(d). For appropriately arranged ensemble of dots the decay 
rates may decrease in such a way that for smaller energy mismatches the final exciton occupation decreases 
with increasing energy mismatch, but after exceeding a critical point it increases. The similar effects in the 
dynamics of single excitons are observed for electron-hole pairs confined in ensembles of four and more dots.

\subsection{Biexciton states}
\label{sec:biexciton}

\begin{figure}[t]
\centering
\includegraphics[scale=0.67]{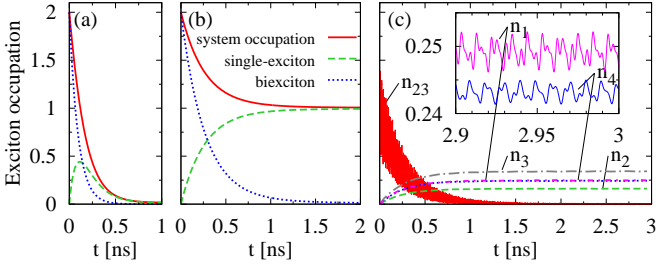}
\caption{Exciton occupation for a biexciton initial state spanned in a system of four QDs 
placed in corners of a square of the side length $r=30$ nm. (a) Superradiant initial state. 
[(b) and (c)] Initial biexciton state which allows for a recombination of only one 
exciton.  
($d_{12}=\sqrt{\Gamma_{33}\Gamma_{44}}$, $d_{13}=-\sqrt{\Gamma_{22}\Gamma_{44}}$,
$d_{14}=\Gamma_{44}\left(\Gamma_{33}-\Gamma_{22}\right)\sqrt{\Gamma_{22}\Gamma_{33}}
/\left(\Gamma_{22}\Gamma_{33}-\Gamma_{11}\Gamma_{44}\right)$,
$d_{23}=-\Gamma_{44}\left(\Gamma_{33}-\Gamma_{22}\right)\sqrt{\Gamma_{11}\Gamma_{44}}
/\left(\Gamma_{22}\Gamma_{33}-\Gamma_{11}\Gamma_{44}\right)$,
$d_{24}=\Gamma_{44}\sqrt{\Gamma_{11}\Gamma_{33}}/\Gamma_{11}$, and 
$d_{34}=-\Gamma_{44}\sqrt{\Gamma_{11}\Gamma_{22}}/\Gamma_{11}$). The values of the decay rates of 
individual dot are the same for all Figs. and take the values: $\Gamma_{11}=2.26$ ns$^{-1}$, 
$\Gamma_{22} = 2.75$ ns$^{-1}$, $\Gamma_{33}=0.88$ ns$^{-1}$, and $\Gamma_{44}=1.5$ ns$^{-1}$.
The biexction shifts are $B_{13}=B_{12}-0.702$ meV, $B_{14}=B_{12}+0.022$ meV, $B_{23}=B_{12}+0.218$ meV, 
$B_{24}=B_{12}-0.741$ meV, and $B_{34}=B_{12}+0.042$ meV with $B_{12}$ being an arbitrary parameter.
The line types used in panel (b) are also valid for panel (a).
}
\label{fig:Biex_fig}
\end{figure}

In multiple QDs built out of three and four units more than one exciton may be delocalized and thus these 
systems allow for more complex collective effect. Below we focus on biexciton states which in general are 
described by a vector $|\mathrm{biexciton}\rangle = \sum^{N}_{i,j, i\neq j}b_{ij}|ij\rangle$, where 
$\sum^{N}_{ij}|b_{ij}|^2=1$. The realistic ensembles of QDs ($\Delta\neq 0$, $\bm{d}_{i}\neq\bm{d}_{j}$, $V_{ij}\neq 0$ and 
$B_{ij}\neq 0$) permit superradiance phenomena in the two-exciton subspace if the biexcitonic, as well as 
single-exciton, eigenstates correspond to the Dicke states. As in the single-exciton case, the biexcitonic 
state is considered to be superradiant if the exciton occupation of a system prepared in this state decays 
totally. Although eigenstates of the Hamiltonian (\ref{hamiltonianS}) do not mix localized basis states 
associated with different exciton numbers, the biexcitonic superradiant state may be formed provided one of 
the single-exciton eigenstates has a superradiant character. Thus the biexcitonic superradiant states occur 
only if collective effects are present in the single-exciton subspace. Due to the equal number of 
single-exciton eigenstates and QDs forming the ensemble the ``collective regime'' in the single-exciton domain 
may be achieved by adjusting only the decay rates (dipole moments). As will be explained in detail below, the 
same rates $\Gamma_{ij}$ (\ref{Gamma}) define the biexcitonic Dicke states. Thus, in order to achieve the 
collective effects in the two-exciton subspace one has to also appropriately adjust the spatial arrangement of 
the dots or energies.

The superradiant two exciton superpositions may be spanned as well in ensembles of four emitters as in 
triple QDs and take a form
\begin{eqnarray}
\label{brightBi}
 |\mathrm{SUPER}\rangle_{\mathrm{B}} = \frac{\sum\limits_{i,j, i\neq j}^{N}\sqrt{\Gamma_{ii}\Gamma_{jj}}|ij\rangle}
 {\sqrt{\sum\limits_{i,j, i\neq j}^{N}\Gamma_{ii}\Gamma_{jj}}}.
\end{eqnarray}
Similarly to the single-exciton case, the biexcitonic superradiant states are defined by the maximum value of 
the transition probability ($\sim|\langle\mathrm{SUPER}|H_{\mathrm{S-rad}}|\mathrm{biexciton}\rangle|^2$), 
but this time from the initial biexciton state to the final single-exciton superradiant one (\ref{bright}).
The form of the condition is governed by the coupling to the radiative environment [Hamiltonian 
(\ref{hamiltonianSrad})] which induces decay of only one exciton at a time. Thus total quenching of two excitons 
must occur through formation of single-exciton superradiant states. As can be seen in Fig. \ref{fig:Biex_fig}(a), 
due to the decay of the biexciton superradiant state a part of initial excitation is initially transferred to the 
single-exciton state which reaches a maximum population and then is totally quenched, together with the biexciton 
state. The condition for the existence of the superradiant state (\ref{brightBi}) implies that the transition to 
subradiant states (\ref{dark}) vanishes.

Two excitons can be blocked in a system if a transition from the biexciton state to any single-exciton state is 
forbidden, i.\ e. the transition matrix element $\langle\mathrm{single}|H_{\mathrm{S-rad}}|\mathrm{biexciton}\rangle$ 
vanishes. Due to infinite number of possible single-exciton states this condition reduces to the requirement 
$H_{\mathrm{S-rad}}|\mathrm{biexciton}\rangle=0$. To simplify the description we define non-normalized 
amplitudes $d_{ij}$ in such a way that the amplitudes of biexciton states $b_{ij}=d_{ij}/\sqrt{\sum_{ij}|d_{ij}|^2}$.
For a triple QD the condition leads to a system of three equations of the form:
\begin{eqnarray*}
  d_{ij}\sqrt{\Gamma_{jj}} + d_{ik}\sqrt{\Gamma_{kk}} = 0, 
  \end{eqnarray*}
where every subscript $i$, $j$, and $k$ takes the values $1$, $2$, and $3$, respectively.
The system is satisfied only in the case of vanishing amplitudes $d_{12}=d_{13}=d_{23}=0$ which means that it is 
impossible to block two excitons in a triple QD. The coefficients of a stable biexcitonic state spanned 
in a system of four QDs must satisfy the system of four equations of the form: 
\begin{eqnarray*}
d_{ij}\sqrt{\Gamma_{jj}} + d_{ik}\sqrt{\Gamma_{kk}} + d_{il}\sqrt{\Gamma_{ll}} = 0,
\end{eqnarray*}
where every subscript $i$, $j$, $k$, and $l$ takes the values $1$, $2$, $3$, and $4$, respectively.
The above equations lead to the condition for the $d_{ij}$ numbers in a form
\begin{eqnarray}
\label{subbiex}
 d_{12} &=& - \frac{\sqrt{\Gamma_{33}}d_{13}+\sqrt{\Gamma_{44}}d_{14}}{\sqrt{\Gamma_{22}}}, \\
 d_{23} &=& \sqrt{\frac{\Gamma_{11}\Gamma_{44}}{\Gamma_{22}\Gamma_{33}}}d_{14},\quad
 d_{24} = \sqrt{\frac{\Gamma_{11}\Gamma_{33}}{\Gamma_{22}\Gamma_{44}}}d_{13} \nonumber \\
 d_{34} &=& - \frac{\sqrt{\Gamma_{11}\Gamma_{33}}d_{13}+\sqrt{\Gamma_{11}\Gamma_{44}}d_{14}}{\sqrt{\Gamma_{33}\Gamma_{44}}}. \nonumber
\end{eqnarray}
Although in a system of four QDs many dark states may be spanned, in realistic systems the parameters of biexciton 
states may be adjusted in such a way that only one, particular (for specified values $d_{13}$ and $d_{14}$) state 
is blocked. Thus, as in the ideal systems and similarly to the double QDs, \cite{sitek12} the contribution of any 
pair of dots to the total biexcitonic population is constant in time ($n_{ij}=\mathrm{const}$).

In realistic QDs one may realize either superradiant 
or subradiant biexciton eigenstate but never both simultaneously as in the ideal system. In both cases 
the basis is supplemented by a third kind of state which allow for a recombination of one electron-hole pair 
and trapping of the second exciton. The evolution of a four QD system prepared in this state is shown in 
Figs. \ref{fig:Biex_fig}(b) and \ref{fig:Biex_fig}(c). While the population of the biexciton state decreases, 
the occupation of single-exciton state increases until it stabilize at the level corresponding to total trapping 
of one electron-hole [green-dashed line \ref{fig:Biex_fig}(b)]. The third basis state must be orthogonal to the 
superradiant state (\ref{brightBi}), which equals to the requirement of vanishing transition matrix element between 
the biexciton state and the single-exciton superradiant state 
($\langle\mathrm{SUPER}|H_{\mathrm{S-rad}}|\mathrm{biexciton}\rangle = 0$) and means that the state 
$H_{\mathrm{S-rad}}|\mathrm{biexciton}\rangle$ has a subradiant character. The orthogonality to the subradiant 
biexcitonic state (\ref{subbiex}) excludes contributions from the two-exciton dark states and thus the population 
trapping occurs only due to formation of a single-exciton subradiant state (\ref{dark}). Since the transition to 
the state (\ref{bright}) and thus the decay of single-exciton states is forbidden, one exciton is blocked in the 
ensemble. Because the biexcitonic subradiant states cannot be formed in triple QDs, the requirement defining 
biexcitonic states which allow to block one exciton reduces to the orthogonality to the superradiant state, and the 
states take a form
\begin{eqnarray*}
 \frac{a_{12}\sqrt{\Gamma_{33}}|12\rangle + a_{13}\sqrt{\Gamma_{22}}|13\rangle + a_{23}\sqrt{\Gamma_{11}}|23\rangle}
 {\sqrt{|a_{12}|^2\Gamma_{33} + |a_{13}|^2\Gamma_{22} + |a_{23}|^2\Gamma_{11}}},
\end{eqnarray*}
where $a_{12}+a_{13}+a_{23}=0$. Whereas, due to the orthogonality to the superradiant state the six amplitudes of 
the biexcitonic states spanned in ensembles of four QDs take a form:
\begin{eqnarray*}
d_{ij} = a_{ij}\sqrt{\Gamma_{kk}\Gamma_{ll}},
\end{eqnarray*}
where the coefficients must satisfy equation $a_{12} + a_{13} + a_{14} + a_{23} + a_{24} + a_{34} = 0$ 
and, using further the orthogonality to the subradiant state relations,
\begin{eqnarray*}
 a_{12}\Gamma_{33}\Gamma_{44} - a_{13}\Gamma_{22}\Gamma_{44} - a_{24}\Gamma_{11}\Gamma_{33} + a_{34}\Gamma_{11}\Gamma_{22} = 0, \\
 a_{12}\Gamma_{33}\Gamma_{44} - a_{14}\Gamma_{22}\Gamma_{33} - a_{23}\Gamma_{11}\Gamma_{44} + a_{34}\Gamma_{11}\Gamma_{22} = 0.
\end{eqnarray*}

Due to the existence of single and biexciton optically inactive states in ensembles of four QDs the initial 
biexcitonic states allow for many combinations of final occupation. An arbitrary fraction of exciton occupation 
($\leq 2$)  may be trapped by an appropriate combination of blocked single-excitons and biexcitons due to 
contribution to the initial state from dark states and basis states which allow to recombine only one 
electron-hole pair. 
Irrespective of the initial number of excitons, ensembles of three QDs allow to span only single-exciton subradiant 
states and thus block only single-excitons.

It is important to emphasis that if the system was prepared in a bright biexcitonic state which leads to trapping 
of single-exciton occupation [Figs. \ref{fig:Biex_fig}(b) and \ref{fig:Biex_fig}(c)], then the pronounced 
oscillations due to coupling between the dots appear only in occupation of localized biexciton states, while the 
amplitudes of oscillations in the populations of single dots are negligible [\ref{fig:Biex_fig}(c)] 
compared to single-exciton initial state [Fig. \ref{fig:3QD_comparison}(d)]. This means that the biexciton initial 
state allows to achieve well defined stable single-exciton  subradiant states, which is important for the application 
of quantum computation.

\subsection{P-i-n junction}
\label{pinjunction}

The presence of optical collective effects in realistic QD systems requires high accuracy of a 
system parameters which may be controlled on the manufacturing stage or by external fields by, 
e.g. implementing the dots into the intrinsic region of a p-i-n junction. This structure provides 
a possibility of a separated injection of electrons and holes into QDs from both sides of a sample 
and control of exciton dynamics and QD parameters through application of contacts on n- and 
p- type regions. It has been shown that controlled with a bias voltage carrier tunneling into 
a single QD in a p-i-n structure incorporated into a microcavity leads to regulated emission of 
single photons and pairs of photons.\cite{Benson00} The ideas were followed by a technological 
realization of an electrically driven single-photon emitter with a layer of self-organized InAs 
QDs.\cite{Yuan02} The gate voltages constructed over dots allow to control energies of the dots 
and dipole moments, but the magnitudes of decay rates of single QDs (\ref{Gamma}) depend on the 
average energy of the ensemble ($E$) and thus operations on one dot change the decay rates 
of all QDs in the system.

\begin{figure}[t]
\centering
\includegraphics[scale=0.7]{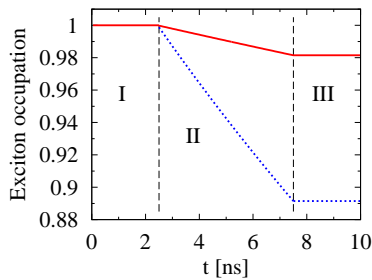}
\caption{Exciton occupation for a system of four QDs placed in vertices of a square of the 
side length $r = 30$ nm and prepared initially in a single-exciton subradiant state.
In regions I and III  the decay rates are the same as in Fig. \ref{fig:Biex_fig}
 i.\ e.\ selected such that the systems interact collectively with the radiative surrounding.  
The initial state is $\sim(\sqrt{\Gamma_{22}\Gamma_{33}\Gamma_{44}}|1\rangle+
\sqrt{\Gamma_{11}\Gamma_{33}\Gamma_{44}}|2\rangle + \sqrt{\Gamma_{11}\Gamma_{22}\Gamma_{44}}|3\rangle
-3 \sqrt{\Gamma_{11}\Gamma_{22}\Gamma_{33}}|4\rangle)$.
The red solid line corresponds to changes of the decay rates of individual QDs, whereas the blue
dotted line to the same changes plus changes of the relative angular orientation of the dipoles.  
In regions II the decay rates and angles have been modified such that the ensemble does not interact 
collectively with its photon environment (see text).  }
\label{fig:pin}
\end{figure}

In order to simulate this possibility we have calculated the time evolution of the system of four QDs and 
we have changed the decay rates at some selected time points.  The results are shown in In Fig. \ref{fig:pin}
where the system is initially in the ``collective regime'' and has been prepared in a single-exciton subradiant 
state.  We assume that the control electric fields are weak enough to exclude ``dissociated'' exciton states.  
As expected, initially the excitation is blocked in the system.  

At time 2.5 ns a change of parameters occurs which destabilizes the system and induces quenching of the excitation.
The effect may be produced by a variation of the internal electric field in the p-i-n junction, which is simulated 
here by a change of the decay rates of each single dot, and shown by the red solid lines in Fig. \ref{fig:pin}. 
The decay rates have been changed as follows: 
$\Gamma_{11}\to 1.4\Gamma_{11}, \
\Gamma_{22}\to 1.3\Gamma_{22},  \
\Gamma_{33}\to 1.2\Gamma_{33},  \ 
\Gamma_{44}\to 1.1\Gamma_{11}.$   
In this case the quenching is relatively weak because of a small change in the ratio of the decay rates but the effect 
is visible. At time 7.5 ns the parameters of each dot are changed back and the system population is again stable.

The decay may be enhanced by changing the orientation of the dipoles as shown by the blue dotted lines in 
Fig. \ref{fig:pin}. All dipoles are parallel in the regions I and III, i.\ e. all angles $\theta_{ij}=0$.
But now, in addition to the previous variations of $\Gamma_{ij}$ the angles are also modified in the regions II: 
$\theta_{12}=0.1, \ \theta_{13}=0.11, \ \theta_{14}=0.2,\ \theta_{23}=0.105, \ \theta_{24}=0.2, \ \theta_{34}=0.1$ 
radians.  The misalignment of the dipole moments creates thus a stronger decay.  
The initial conditions may be restored at any time which may result in the trapping of a desired fraction of the 
initial occupation.

\section{Conclusions}
\label{sec:conclusions}

We have studied the optical collective effects due to interaction of multiple quantum dots built of 
three and four emitters with the radiative surrounding. Ensembles of three and more emitters allow 
to span many subradiant states which facilitate preparation of the system in an optically inactive
single-exciton states and for ensemble of four emitters also in the biexciton subspace.

We specified the conditions which allow the superradiance phenomena to occur in coupled inhomogeneous 
systems with different fundamental transition energies and dipole moments (and thus decay rates). 
We discussed the dynamics of single electron-hole pairs and biexcitons. Although many features typical 
for identical atoms, such as spontaneous trapping of excitation, may also occur in inhomogeneous QDs 
there are differences in the dynamics of these systems. In principle, coupling between the dots induces 
excitation transfer between the dots which together with a possibility to define many dark states in 
ensembles of three and more dots lead to oscillations in the occupation of single dots. The amplitudes of
these oscillates may be considerably reduced if the system is prepared initially in an appropriate biexction 
state which allows for trapping of one electron-hole pair.

We envision that the presented collective effects may be controlled if the ensemble of dots is 
placed in the intrinsic region of a p-i-n junction with contacts constructed over the dots which
due to sufficiently weak electric fields allow to control the dynamics of excitons and thus 
 collective effects.


\begin{acknowledgments}
This work was funded by the Icelandic Research Fund (Rannis) and by a 
Polish NCN grant No. DEC-2011/01/B/ST3/02415.
A. S. acknowledges support within a scholarship for outstanding young
scientists granted by the Polish MNiSW. The authors are grateful to 
Pawe{\l} Machnikowski for fruitful discussions. 
\end{acknowledgments}

\appendix*
\section{Derivation of the equation (\ref{single})}

To express an arbitrary single-exciton state in terms of the superradiant state  (\ref{bright}) 
and a subradiant state (\ref{dark}) we begin with a derivation of a formula for a localized state 
$|i\rangle$. The subradiant state which allows to cancel out all different from the state $|i\rangle$ 
localized contributions to the superradiant state has a form
\begin{eqnarray}
\label{SUBi}
 |\mathrm{SUB}\rangle_{i} = \frac{\sum\limits^{N}_{j\neq i}\Gamma_{jj}|i\rangle
 -\sqrt{\Gamma_{ii}}\sum\limits^{N}_{j\neq i}\sqrt{\Gamma_{jj}}|j\rangle}
 {\sqrt{\sum\limits^{N}_{j\neq i}\Gamma_{jj}\sum\limits^{N}_{i = 1}\Gamma_{ii}}}
\end{eqnarray}
and is orthogonal to the superradiant state for an arbitrarily chosen state $|i\rangle$. Therefore the 
localized single-exciton states may be decomposed into a superposition of the superradiant (\ref{bright}) 
and subradiant state defined in the formula (\ref{SUBi}) according to the equation
\begin{eqnarray}
\label{statei}
 |i\rangle = \frac{\sqrt{\Gamma_{ii}}|\mathrm{SUPER}\rangle + \sqrt{\sum\limits^{N}_{k\neq i}\Gamma_{kk}}|\mathrm{SUB}\rangle_{i}}
 {\sqrt{\sum\limits^{N}_{i = 1}\Gamma_{ii}}}.
\end{eqnarray}
The above formula allows to define a sub- and superradiant component in every single-exciton state 
\begin{eqnarray*}
 |\mathrm{s}\rangle = \sum\limits^{N}_{i = 1}c_{i}|i\rangle = |\mathrm{SUPER}\rangle' + |\mathrm{SUB}\rangle'.
\end{eqnarray*}
Since there is only one superradiant state in any system of $N$ QDs, the short-living contribution  
\begin{eqnarray*}
   |\mathrm{SUPER}\rangle' = \frac{\sum\limits^{N}_{i=1}c_{i}\sqrt{\Gamma_{ii}}
 }{\sqrt{\sum\limits^{N}_{i=1}\Gamma_{ii}}}|\mathrm{SUPER}\rangle
\end{eqnarray*}
is proportional to the state (\ref{bright}) with a weight factor 
$\left(\sum\limits^{N}_{i=1}c_{i}\sqrt{\Gamma_{ii}}\right)
/\sqrt{\sum\limits^{N}_{i=1}\Gamma_{ii}}$, while the stable part
\begin{eqnarray*}
   |\mathrm{SUB}\rangle' &=& \frac{\sum\limits^{N}_{i=1}c_{i}\sqrt{\sum\limits^{N}_{k\neq i}\Gamma_{kk}}}
   {\sqrt{\sum\limits^{N}_{i = 1}\Gamma_{ii}}}|\mathrm{SUB}\rangle_{i} \\
   &=& \frac{\sum\limits^{N}_{i = 1}c_{i}\left(\sum\limits^{N}_{j\neq i}\Gamma_{jj}|i\rangle 
   - \sum\limits^{N}_{j\neq i}\sqrt{\Gamma_{ii}\Gamma_{jj}}|j\rangle\right)}{\sum\limits^{N}_{i = 1}\Gamma_{ii}}
\end{eqnarray*}
is a combination of $N$ subradiant states (\ref{SUBi}), which as a sum of dark states remains 
optically inactive irrespective of a number of emitters. The component $|\mathrm{SUB}\rangle'$ is 
proportional to the subradiant state defined by the formula (\ref{dark2}) with an amplitude 
$\sqrt{1-\left(\sum\limits^{N}_{i=1}c_{i}\sqrt{\Gamma_{ii}}/\sqrt{\sum\limits^{N}_{i=1}\Gamma_{ii}}\right)^2}$.


\end{document}